\documentstyle[12pt]{article}
\newcommand\bq{\begin{quotation}}
\newcommand\eq{\end{quotation}}
\newcommand\sig{\bf\sigma}
\begin{document}
\title{Bohmian Mechanics and the Quantum Revolution}

\author{Sheldon Goldstein\\
Department of Mathematics, Rutgers University\\ New Brunswick, NJ
08903}
\date{(March 31, 1995)}
\maketitle

\section*{The Quantum Revolution: An Orthodox Perspective}
\openup-.05\jot

When I was young I was fascinated by the quantum revolution: the transition
{}from classical definiteness and determinism to quantum indeterminacy and
uncertainty, from classical laws that are indifferent, if not hostile, to
the human presence, to quantum laws that fundamentally depend upon an
observer for their very meaning. I was intrigued by the radical
subjectivity, as expressed by Heisenberg's assertion \cite{Heisenberg} that
``The idea of an objective real world whose smallest parts exist
objectively in the same sense as stones or trees exist, independently of
whether or not we observe them \dots\  is impossible \dots''  It is true that I
did not really understand what the quantum side of this transition in fact
entailed, but that very fact made quantum mechanics seem to me all the more
exciting. I was eager to learn precisely what the alluring quantum
mysteries might mean, what kind of world they describe, as well as exactly
what evidence could compel---or at least support---such radical
conclusions.

The subjectivity and indeterminism of quantum theory were accompanied by
the view that quantum particles don't really have trajectories and that, by
Heisenberg's uncertainty principle, any talk of such things is meaningless.
While the conclusion of meaninglessness seemed a bit forced, I was
reassured by one of the best quantum mechanics textbooks, that of Landau
and Lifshitz, which declared flatly in referring to the double-slit
experiment that \cite{LL} ``It is clear that this result can in no way be
reconciled with the idea that electrons move in paths. \dots\ In quantum
mechanics there is no such concept as the path of a particle.''

I was, however, given pause by the fact that strong disapproval of the
accepted understanding of the quantum revolution was expressed by a few
physicists, including at least two of the founders of quantum theory:
Schr\"odinger, the creator of wave mechanics, who maintained that ``Bohr's
\dots\  approach to atomic problems \dots\   is really remarkable. He is
completely
convinced that any understanding in the usual sense of the word is
impossible'' (letter to Wien, quoted in \cite{Moore}); and Einstein
\cite{Einstein}, who believed that ``the essentially statistical character
of contemporary quantum theory is solely to be ascribed to the fact that
this (theory) operates with an incomplete description of physical
systems.''

But I was instructed that Einstein's belief was out of the question,
since John von Neumann had proven, with the utmost mathematical rigor, that
a return by physics to any sort of fundamental determinism was impossible.
Von Neumann himself \cite{vN} concluded that ``It is therefore not, as is
often assumed, a question of a re-interpretation of quantum mechanics---the
present system of quantum mechanics would have to be objectively false, in
order that another description of the elementary processes than the
statistical one be possible.'' According to Max Born, the physicist who
formulated the now standard statistical interpretation of the wave
function, von Neumann had shown that \cite{born} ``no concealed parameters
can be introduced with the help of which the indeterministic description
could be transformed into a deterministic one.  Hence if a future theory
should be deterministic, it cannot be a modification of the present one but
must be essentially different. How this could be possible without
sacrificing a whole treasure of well established results I leave to the
determinists to worry about.''

Nonetheless, I did hear rumors now and then about a certain
hidden variables theory constructed by David Bohm in 1952 \cite{Bohm52}, a
theory that was reputed somehow to recast quantum theory into a completely
deterministic form. I was not at all clear about what exactly Bohm had
proposed, but I was assured by my teachers that whatever it was, it could
not possibly accomplish what Bohm had claimed.

Occasionally I would hear that there might be some flaw in the proof of von
Neumann, but I was advised not to worry about this because there were
other, {\it even more convincing\/}, proofs of von Neumann's result! What
an embarrassment of riches! For example, according to Eugene Wigner \cite[page
291]{Wigner1} ``the proof he [von Neumann] published \dots, though it was
made much more convincing later on by Kochen and Specker, still uses
assumptions which, in my opinion, can quite reasonably be questioned. \dots\
In my opinion, the most convincing argument against the theory of hidden
variables was presented by J. S. Bell (1964).'' Elsewhere \cite{Wigner2}
Wigner reports, concerning Bohm's theory, that ``This is an interesting
idea and even though few of us were ready to accept it, it must be admitted
that the truly telling argument against it was produced as late as 1965, by
J. S.  Bell. \dots\  This appears to give a convincing argument against the
hidden variables theory.''

Were I young today---and wished to be reassured about the health of the
quantum revolution---I could turn for support to Murray Gell-Mann, who
reports \cite{GM} that ``for more than forty years, David [Bohm] tried to
reformulate and reinterpret quantum mechanics so as to overcome his
doubts,'' adding that ``some theoretical work of John Bell revealed that
the EPRB experimental setup could be used to distinguish quantum mechanics
{}from hypothetical hidden variable theories \dots\  After the publication of
Bell's work, various teams of experimental physicists carried out the EPRB
experiment. The result was eagerly awaited, although virtually all
physicists were betting on the correctness of quantum mechanics, which was,
in fact, vindicated by the outcome.''

In view of what I've so far said, it would be natural to suppose that the
two books with which I shall be concerned in this essay, {\it The Undivided
Universe\/} by Bohm and Basil J. Hiley, which is a presentation of Bohm's
1952 theory incorporating some later developments, many of which represent
joint work with Bohm's long-time collaborator Hiley, and Bell's {\it
Speakable and Unspeakable in Quantum Mechanics\/}, which collects all of
Bell's work, up to 1987, on the foundations of quantum mechanics, are in
diametric opposition, with the former defending hidden variables and the
latter defending quantum orthodoxy. This, however, is not so; in fact, Bell
provides a far stronger argument against the orthodox Copenhagen
interpretation of quantum mechanics and, indeed, in favor of what have been
called (absurdly, according to Bell) hidden variables theories than do Bohm
and Hiley!

It might thus seem, and it has in fact been suggested \cite{Speiser}, that
there must have been two Bells---that Bell must have been
schizophrenic. However, there was, unfortunately for us, but one Bell, and
he was one of the sanest and most rational of men. Bell did not---as did in
fact Bohm---at some point convert from a defender to a critic of quantum
orthodoxy; he was always unhappy with traditional quantum mechanics, and
reports \cite{Bernstein} that even as a student learning quantum theory ``I
hesitated to think it might be wrong, but I {\it knew\/} that it was
rotten.''

But what, you might well ask, of the statements of Wigner and Gell-Mann, in
effect praising Bell for destroying Bohm?  The best that can be said for
these statements, by two of the leading physicists of their respective
generations, is that they are grossly misleading, since Bell's analysis is
entirely compatible with Bohm's theory and in fact affords this theory
dramatically compelling support!

\section*{Bohmian Mechanics}

What Bohm did in 1952---the ``proof'' of von Neumann and the claims of
Born, Bohr, Landau and Lifshitz, and Heisenberg notwithstanding---was to
provide an objective and completely deterministic account of
nonrelativistic quantum phenomena through a refinement and completion of de
Broglie's 1926 pilot-wave model.  In Bohm's theory, which he sometimes called
the causal interpretation of quantum mechanics and sometimes the theory of
the quantum potential but which is nowadays more frequently called Bohmian
mechanics,  a system of particles is described in part by its wave
function, evolving, as usual, according to Schr\"odinger's equation;
however, the wave function provides only a partial description of the
system, which is completed by the specification of the actual positions of
the particles. The latter evolve according to Bohm's ``guiding equation,''
which expresses the velocities of the particles in terms of the wave
function.  Thus, in Bohmian mechanics the configuration of a system of
particles evolves via a deterministic motion choreographed by the wave
function. In particular, when a particle is sent into a double-slit
apparatus, the slit through which it passes and where it arrives on the
photographic plate are completely determined by its initial position and
wave function.

Bohm's achievement was greeted by most physicists with indifference, if not
with outright hostility.  Bell's reaction, however, was a refreshing exception:

\bq\noindent In 1952 I saw the impossible done.  It was in papers by David
Bohm. Bohm showed explicitly how parameters could indeed be introduced,
into nonrelativistic wave mechanics, with the help of which the
indeterministic description could be transformed into a deterministic one.
More importantly, in my opinion, the subjectivity of the orthodox version,
the necessary reference to the `observer,' could be eliminated.  \dots\  But
why then had Born not told me of this `pilot wave'?  If only to point out
what was wrong with it? Why did von Neumann not consider it? \dots\  Why is the
pilot wave picture ignored in text books?  Should it not be taught, not as
the only way, but as an antidote to the prevailing complacency? To show us
that vagueness, subjectivity, and indeterminism, are not forced on us by
experimental facts, but by deliberate theoretical choice?'' (Bell, page
160)
\eq
and
\bq\noindent Bohm's 1952 papers on quantum mechanics were for me a
revelation. The elimination of indeterminism was very striking. But more
important, it seemed to me, was the elimination of any need for a vague
division of the world into ``system'' on the one hand, and ``apparatus'' or
``observer'' on the other. I have always felt since that people who have
not grasped the ideas of those papers\dots\  and unfortunately they remain the
majority \dots\  are handicapped in any discussion of the meaning of quantum
mechanics. (Bell, page 173)
\eq

In view of what has so often been said---by most of the leading physicists
of this century and in the strongest possible terms---about the radical
implications of quantum theory, it is not easy to accept that Bohmian
mechanics really works.  However, in fact, it does: Bohmian mechanics
accounts for all of the phenomena governed by nonrelativistic quantum
mechanics, from spectral lines and quantum interference experiments to
scattering theory and superconductivity.  In particular, the usual
measurement postulates of quantum theory, including collapse of the wave
function and probabilities given by the absolute square of probability
amplitudes, emerge as a consequence merely of the two equations of motion
for Bohmian mechanics---Schr\"odinger's equation and the guiding
equation---without the traditional invocation of a special and somewhat
obscure status for observation.

Both of these books provide explanations of how this comes about. You will
not find in these books the evasions---by now almost standard fare for the
foundations of quantum mechanics---in which the real problems are skirted
rather than solved.  The account of Bohm and Hiley is quite detailed;
Bell's is rather brief. The presentation of Bohm and Hiley is well written and
insightfully describes a broad spectrum of important physics. Bell's
is almost always precisely on target and almost never less than
thoroughly masterful. Whereas Bohm and Hiley provide {\it an\/}
explanation, it is Bell that supplies {\it the \/} explanation.  Bell's
presentation affords by far the deeper appreciation of the essence of
Bohmian mechanics, of its virtues and of its limitations.

\section*{Nonlocality}

One of the most remarkable implications of quantum theory---a {\it
genuine\/} implication, unlike those which intrigued me when I was younger
and which reflect the prejudices of their proponents far more than they do
anything about the demands of nature---is that of quantum nonlocality. It
is to this that the main title of the book of Bohm and Hiley refers. No
physicists have contributed as much to our understanding of quantum
nonlocality as have Bell and Bohm. This subject plays a central role in
every article in Bell's book, and Bohm and Hiley also give it considerable
attention. Both treatments of this subject are excellent, with Bell's
nearly perfect.

A great many physicists and philosophers have regarded nonlocality as
scientifically and philosophically unacceptable. In this regard Bohm and
Hiley make an observation with which it would be difficult to disagree:

\bq For several centuries, there has been a strong feeling that nonlocal
theories are not acceptable in physics. It is well known, for example, that
Newton felt very uneasy about action-at-a-distance and that Einstein
regarded it as `spooky'. One can understand this feeling, but if one
reflects deeply and seriously on this subject one can see nothing basically
irrational about such an idea. Rather it seems to be most reasonable to
keep an open mind on the subject and therefore to allow oneself to explore
this possibility. If the price of avoiding nonlocality is to make an
intuitive explanation  impossible, one has to ask whether the cost is not
too great. (BH, page 57)
\eq
However, in linking nonlocality with the desire for an ``intuitive
explanation,'' this observation of Bohm and Hiley is somewhat inadequate.
The extent of this inadequacy can best be appreciated by way of comparison with
Bell's  treatment of nonlocality.

First of all, Bell observes that quantum nonlocality is implicit in the
fundamental mathematical structure of quantum mechanics, as provided by a
wave function on the configuration space of a many-particle system. Thus,
since the very same wave function is also central to Bohmian mechanics, we
should not be surprised to find, as indeed we do, that it, too, is
nonlocal. Bohmian mechanics, however, is {\it explicitly\/} nonlocal while the
nonlocality of orthodox quantum theory is ambiguous and merely
implicit---or so it was until 1964, when Bell showed that the very
predictions of standard quantum theory provide unmistakable, though
indirect, evidence for this nonlocality. As Bell has emphasized:

\bq That the guiding wave, in the general case, propagates not in ordinary
three-space but in a multidimensional-configuration space is the origin of
the notorious `nonlocality' of quantum mechanics. It is a merit of the de
Broglie-Bohm version to bring this out so explicitly that it cannot be
ignored. (Bell, page 115)
\eq

In the last quoted sentence, Bell is referring to his own analysis of
Bohmian mechanics, some fifteen years earlier, in the course of which he
observed that

\bq \noindent
\dots\  in this theory an explicit causal mechanism exists whereby the
disposition of one piece of apparatus affects the results obtained with a
distant piece.

Bohm of course was well aware of these features of his scheme, and has
given them much attention. However, it must be stressed that, to the
present writer's knowledge, there is no {\it proof\/} that {\it any\/}
hidden variable account of quantum mechanics {\it must\/} have this
extraordinary character. It would therefore be interesting, perhaps, to
pursue some further ``impossibility proofs,'' replacing the arbitrary
axioms objected to above by some condition of locality, or of separability
of distant systems. (Bell, page 11)
\eq
In a footnote, Bell added that ``Since the completion of this paper such a
proof has been found.'' He is referring, of course, to his celebrated paper
\cite{Bell1}
``On the Einstein-Podolsky-Rosen Paradox,'' in which he derives Bell's
inequality---on the basis of which he concludes that the predictions of quantum
theory are irreducibly nonlocal.

Thus Bell's refutation, referred to by Wigner and Gell-Mann, of the theory
of hidden variables is directed only against local hidden variables and
does not touch Bohmian mechanics, which is nonlocal.  It is true that if
the nonlocality of Bohmian mechanics had not already been noticed, Bell's
theorem could have been used to draw the conclusion that Bohmian mechanics
must be nonlocal, and thus, for some, unacceptable. But Bohmian mechanics
is explicitly nonlocal and is not rendered any more nonlocal---or any less
acceptable---because of Bell's theorem.  On the contrary, as the passage
{}from which I have just quoted demonstrates, Bell's recognition of this
feature of Bohmian mechanics is the origin of Bell's inequality, which
shows that there exists no hidden variables theory that improves upon
Bohmian mechanics by avoiding its nonlocality.

Moreover, as I have already indicated, Bell's analysis shows much more. It
shows not only that any hidden variables account of quantum phenomena must
be nonlocal, but that nonlocality is implied merely by the observational
consequences of standard quantum theory itself, so that if nature is
governed by these predictions, then nature is nonlocal!  That nature is so
governed, even in the crucial EPR-correlation experiments, has by now been
established by a great many experiments, the most conclusive of which is
perhaps that of Aspect \cite{Aspect}.  Thus the statement in which Bohm and
Hiley describe what is to be concluded from the results of Aspect's
experiment, namely that ``we have an experimental proof that if there are
hidden variables they must be nonlocal'' (BH, page 144), does not go far
enough.

It must be admitted that on this rather crucial point even Bell's writing
was not always quite as clear as it needed to be.  It would therefore be
worthwhile to spend some time tracing the evolution of, not so much his
ideas as, his mode of expression on this matter.  In the Bell's inequality
paper itself (1964) we find Bell saying that

\bq \noindent It is the requirement of locality, or more precisely that the
result of a measurement on one system be unaffected by operations on a
distant system with which it has interacted in the past, that creates the
essential difficulty. There have been attempts to show that even without
such a separability or locality requirement no `hidden variable'
interpretation of quantum mechanics is possible. These attempts have been
examined elsewhere and found wanting. Moreover, a hidden variable
interpretation of elementary quantum theory has been explicitly
constructed. That particular interpretation has indeed a grossly non-local
structure. This is characteristic, according to the result to be proved
here, of any such theory which produces exactly the quantum mechanical
predictions. (Bell, page 14)
\eq
However, Bell's argument in this paper demonstrates the stronger
conclusion. In fact, this is clear from his very next paragraph, in which
Bell summarizes the argument of Einstein, Podolsky, and Rosen to the effect
that the assumption of locality implies the existence of certain elements
of reality or, what amounts to more or less the same thing, local hidden
variables or what Bell calls here ``predetermined'' results:

\bq \noindent Consider a pair of spin one-half particles formed somehow in
the singlet state and moving freely in opposite directions. Measurements
can be made, say by Stern-Gerlach magnets, on selected components of the
spins $\sig_1$ and   $\sig_2$. If measurement of the component
$\sig_1\cdot\bf a,$ where $\bf a$ is some unit vector, yields the value
$+1$ then, according to quantum mechanics,  measurement of
$\sig_2\cdot\bf a$ must yield the value $-1$ and vice versa. Now we make
the hypothesis, and it seems one at least worth considering, that if the
two  measurements are made at places remote from one another the orientation
of one magnet does not influence the result obtained with the other. Since
we can predict in advance the result of measuring any chosen component of
$\sig_2$, by previously measuring the same component of $\sig_1$, it
follows that the result of any such measurement must actually be
predetermined. (Bell, page 15)
\eq
Bell then proceeds to show that these EPR elements of reality must satisfy
a certain (Bell's) inequality incompatible with the predictions of quantum
theory.

In 1981 Bell is more explicit
\bq
Could it be that this strange non-locality is a peculiarity of the very
particular de Broglie-Bohm construction \dots\  and could be removed by a more
clever construction? I think not. It now seems that the  non-locality is
deeply rooted in quantum mechanics itself and will persist in any
completion. (Bell, page 132)
\eq
and, indeed, quite emphatic
\bq It is important to note that to the limited degree to which {\it
determinism\/} plays a role in the EPR argument, it is not assumed but
{\it inferred\/}. What is held sacred is the principle of `local
causality'---or `no action at a distance'\dots

It is remarkably difficult to get this point across, that determinism is
not a {\it presupposition\/} of the analysis. (Bell, page 143)
\eq
\bq Let me summarize once again the logic that leads to the impasse. The
EPRB correlations are such that the result of the experiment on one side
immediately foretells that on the other, whenever the analyzers happen to
be parallel. If we do not accept the intervention on one side as a causal
influence on the other, we seem obliged to admit that the results on both
sides are determined in advance anyway, independently of the intervention
on the other side, by signals from the source and by the local magnet
setting. But this has implications for non-parallel settings which conflict
with those of quantum mechanics. So we {\it cannot\/} dismiss intervention
on one side as a causal influence on the other. (Bell, page 149)
\eq
\bq  \dots\  Despite my insistence that the determinism was inferred
rather than assumed, you might still suspect somehow that it is a
preoccupation with determinism that creates the problem. Note well then
that the following argument makes no mention whatever of determinism. \dots\
Finally you might suspect that the very notion of particle, and particle
orbit \dots\  has somehow led us astray\dots. So the following argument
will not
mention particles \dots\  nor any other picture of what goes on at the
microscopic level. Nor will it involve any use of the words `quantum
mechanical system', which can have an unfortunate effect on the discussion.
The difficulty is not created by any such picture or any such terminology.
It is created by the predictions about the correlations in the visible
outputs of certain conceivable experimental set-ups. (Bell, page 150)
\eq

Later still,  Bell identifies the characteristic  feature of any formulation of
quantum mechanics with  manifest nonlocality  as merely
clarity and precision. Bell 1984:

\bq The de Broglie-Bohm picture disposes of the necessity to divide the
world somehow into system and apparatus. But another problem is brought
into focus. This picture, and indeed, I think, any sharp formulation of
quantum mechanics, has a very surprising feature: the consequences of
events at one place propagate to other places faster than light. (Bell,
page 171)
\eq
And 1986:
\bq The very clarity of this picture puts in evidence the extraordinary
`non-locality' of quantum theory. (Bell, page 194)
\eq

\medskip

\section*{Quantum Observables and the Impossibility of Hidden Variables}

So much for what Wigner regarded as ``the truly telling argument against''
the possibility of hidden variables. What about the other ``proofs'' of the
impossibility of hidden variables---for example, those of von Neumann
\cite{vN}, of Gleason \cite{Gleason}, of Kochen and Specker \cite{KS}, and of
Jauch and Piron \cite{JP}---having little if anything to do with
nonlocality? It should not be necessary to mention that the existence of
Bohmian mechanics conclusively demonstrates that none of these could
possibly prove any such thing. It is, however, mildly astonishing that
Bohmian mechanics preceded all of these ``proofs,'' except for von
Neumann's, by at least six years!

Be that as it may, the question arises as to just where these proofs go
wrong. It is with precisely this question that Bell begins his analysis of
the problem of hidden variables in quantum mechanics, in a paper
\cite{Bell2} in the {\it Reviews of Modern Physics\/} written before his
EPR paper (1964) but published only in 1966:

\bq The realization that von Neumann's proof is of
limited relevance has been gaining ground since the 1952 work of Bohm.
However, it is far from universal. Moreover, the writer has not found in
the literature any adequate analysis of what went wrong. Like all authors
of noncommissioned reviews, he thinks that he can restate the position with
such clarity and simplicity that all previous discussions will be eclipsed.
(Bell, page 2)
\eq
In fact, Bohm himself had addressed this very question in his 1952 hidden
variables paper, a fact of which Bell was well aware when he complained
of the inadequacy of the literature on this subject. In a
footnote on this sentence Bell  informs us that

\bq\noindent In particular the analysis of Bohm seems to lack clarity, or else
accuracy. He fully emphasizes the role of the experimental arrangement.
However, it seems to be implied \dots\  that the circumvention of the theorem
{\it requires\/}  the association of {\it hidden\/} variables with the
apparatus as well as with the system observed. The scheme of Section 2
is a counter example to this. Moreover, it will be seen in Section 3 that
if the essential additivity assumption of von Neumann were granted, hidden
variables wherever located would not avail. (Bell, page 12)
\eq
Bell's criticism of Bohm's analysis is completely on target. It is therefore
somewhat unfortunate that some twenty seven years later we find Bohm and Hiley
still insisting that

\bq \noindent\dots\  the essential point is that von Neumann had in mind hidden
parameters that belonged only to the observed system itself and were not
affected by the apparatus. (BH, page 118)
\eq

As Bell said, what was essential for von Neumann's argument {\it per se\/}
was his linearity assumption, not the location of the hidden parameters.
However, we need not go into this linearity assumption here because
Gleason, Bell, and Kochen and Specker have shown that it is, in fact,
completely unnecessary, that (for a Hilbert space of dimension at least 3)
whatever can be proven concerning the impossibility of hidden variables by
invoking this assumption can, by a different argument, also be proven
without it. Rather, the critical assumption behind just about all of the
proofs of the impossibility of hidden variables (except, it might be
argued, for those revolving around locality) is that of {\it
noncontextuality\/}, ``that measurement of an observable must yield the
same value independently of what other measurements may be made
simultaneously'' rather than depending upon ``the complete disposition of
the apparatus'' (Bell, page 9).

Here Bell puts his finger on the essential point. However, just as happened
with his discussion of nonlocality, Bell did not express himself at first
quite as sharply as he did later. The crucial thing to observe is that,
like any good physicist, Bell here speaks without embarrassment of the
measurement of a quantum observable; that is, he does not feel compelled,
as he later did, to surround the word ``measurement'' with quotation marks
in such situations. For example, in 1982 Bell explains that

\bq \noindent the extra assumption is this: the result of `measuring' $P_1$
is independent of which complementary set, $P_2\dots$ or  $P'_2\dots,$ is
`measured' at the same time. \dots\  We are doing a different experiment when
we arrange to  `measure' $P'_2\dots$ rather than $P_2\dots$  along with
$P_1$. (Bell, page 165)
\eq
Moreover, the reason that we find, some seventeen years later, Bell writing in
this way is not that he had become more sensitive to grammatical niceties
in the intervening years. Rather it reflected his frustration with the
continuing abuse of the word ``measurement'' in discussions on the
foundations of  quantum mechanics and the confusion that this abuse makes
all too common, as well as his desire not himself to contribute to this
confusion. As Bell goes on to say

\bq A final moral concerns terminology. Why did such serious people take so
seriously axioms which now seem so arbitrary? I suspect that they were
misled by the pernicious misuse of the word `measurement' in contemporary
theory. This word very strongly suggests the ascertaining of some
preexisting property of some thing, any instrument involved playing a
purely passive role. Quantum experiments are just not like that, as we
learned especially from Bohr. The results have to be regarded as the joint
product of `system' and `apparatus,' the complete experimental set-up. But
the misuse of the word `measurement' makes it easy to forget this and then
to expect that the `results of measurements' should obey some simple logic
in which the apparatus is not mentioned. The resulting difficulties soon
show that any such logic is not ordinary logic.  It is my impression that
the whole vast subject of `Quantum Logic' has arisen in this way from the
misuse of a word. I am convinced that the word `measurement' has now been
so abused that the field would be significantly advanced by banning its use
altogether, in favour for example of the word `experiment.' (Bell, page 166)
\eq

Now Bohm and Hiley also emphasize the importance of contextuality, but they
do not get the emphasis quite right. They say, for example, that ``The
context dependence of results of measurements is a further indication of
how our interpretation does not imply a simple return to the basic
principles of classical physics.'' Maybe so. But the context dependence of
the results of measurements is quite a different matter from the context
dependence of the results of ``measurements.'' By insisting upon speaking
of measurements of quantum observables in the usual careless way deplored
by Bell, by not taking Bell's admonition to heart, they make contextuality
seem a much more striking innovation than  would be expected to be
supported by the rather unremarkable observation that the result of an
experiment should depend upon how it is performed.

It is, however, somewhat curious that what are otherwise rather different
experiments should have come to be regarded by quantum physicists as
``measurements'' of the same ``observable.'' For an analysis of how, from
the perspective of Bohmian mechanics, experiments are naturally associated
with operators---the quantum observables---and in a many-to-one manner, see
\cite{DDGZ}. I note here merely that any such analysis must take into
account Bell's admonition that \bq\noindent \dots\ in physics the only
observations we must consider are position observations, if only the
positions of instrument pointers. It is a great merit of the de
Broglie-Bohm picture to force us to consider this fact. If you make axioms,
rather than definitions and theorems, about the `measurement' of anything
else, then you commit redundancy and risk inconsistency. (Bell, page 166)
\eq

\section*{A Matter of Ontology}
Bohm and Hiley call Bohmian mechanics an ontological interpretation of
quantum theory and in so doing they underline a crucial aspect of the
theory. What they have in mind is that Bohmian mechanics is grounded in
ontology---given by particles described by their positions in space for the
nonrelativistic theory---rather than, like the Copenhagen interpretation, in
epistemology. Bohmian mechanics directly governs the behavior of the basic
elements of this ontology and it is out of this behavior that the
observational principles expressed by the quantum formalism emerge.
However, the essential point is not {\it some\/} ontology versus {\it no\/}
ontology but a {\it clear\/} ontology versus a {\it vague\/} ontology.
After all, one would imagine that any physical theory must at {\it some\/}
level invoke {\it some\/} ontology, and, in fact, the Copenhagen
interpretation presupposes the classical ontology (and it would seem from
what is so often written even classical physics) on the macroscopic
level---without which, so the argument goes, there could be no coherent
discussion of the results of measurement.  For the Copenhagen
interpretation there is, of course, no ontology for the microscopic level,
but would this be so very bad were the notion of the macroscopic not so
vague? In surveying the major possibilities for the interpretation of
quantum mechanics, Bell noted concerning one of these that

\bq  \noindent \dots\  it may  be that Bohr's intuition was right---in that
there is no reality below some `classical' `macroscopic' level. Then
fundamental physical theory would remain fundamentally vague, until
concepts like `macroscopic' could be made sharper than they are today.
(Bell, page 155)
\eq

\section*{The Quantum Potential and the Structure of Bohmian Mechanics}

A reference to the quantum potential, a concept that Bohm and Hiley present
as marking the departure of Bohmian mechanics from classical physics, is
nowhere to be found in Bell's book. This not because Bell didn't like the
expression. It was the concept itself, however expressed, that Bell did not
like, and it was the concept itself, and not just the expression, that he
completely avoided.  This might make it difficult  to appreciate
that the two books under consideration here are concerned with the same
physical theory, since one of them eschews what appears to be the central
innovation of the theory described by the other.

Bohm and Hiley, following Bohm's 1952 paper \cite{Bohm52}, arrive at
Bohmian mechanics by first writing the wave function $\psi$ in the polar
form $\psi=Re^{iS/\hbar}$ where $S$ is real and $R\ge 0$. They then rewrite
Schr\"odinger's equation in terms of these new variables, obtaining a pair
of coupled evolution equations. In particular, they obtain a ``continuity
equation'' for $\rho=R^2$, which, as usual, suggests that $\rho$ be
interpreted as a probability density, and a modified Hamilton-Jacobi
equation for $S$, differing from the usual classical Hamilton-Jacobi
equation only by the appearance of an extra term, the quantum potential,
alongside the classical potential energy term.

Bohm and Hiley then use the modified Hamilton-Jacobi equation to define
particle trajectories just as is done for the classical Hamilton-Jacobi
equation. The resulting motion is precisely what would have been obtained
classically if the particles were acted upon by the force generated by the
quantum potential in addition to the usual forces. Bohm and Hiley point out
that the quantum potential has many unusual properties, which, they say,
account for all the quantum miracles, from the two-slit experiment to
nonlocality.

The quantum potential is a function on configuration space which is
determined by the wave function---it is a functional of the wave
function---whose detailed form (which happens to be
$-\frac{\hbar^2}{2m}\frac{\nabla^2R}{R}$) need not concern us here. It is,
however, worth noting that the quantum potential does depend upon $R$ (and
only on $R$). This seems to suggest that Bohmian mechanics requires a
peculiar dependence of dynamics upon probability and naturally leads to
Wigner's question \cite[page 290]{Wigner1} as to ``how the properties of
the ensemble, described by $R^2$, can influence the motion of an individual
system, the system which $S$ is supposed to describe.''

Now Bohm and Hiley were well aware of such objections, and carefully
explain why they are not valid---as did Bohm already in 1952. The essential
point is that the identification of $R^2$ with probability is somehow of a
secondary character, and that the fundamental role of $R$, as well as of $S$,
is the dynamical one. Nonetheless, formulating Bohmian mechanics directly
in terms of variables that we naturally regard as having a probablistic
character is an invitation to such objections, an invitation that most
physicists are glad to accept!

It must be admitted that the problem of the status of probability in
Bohmian mechanics---of the identification of ${|\psi|}^2$ with the
probability density for position---is a subtle one, and that the analysis
of Bohm and Hiley of this issue is neither entirely adequate nor completely
up to date. (For a recent detailed treatment, see \cite{DGZ}.) However,
whatever inadequacies exist in the treatment of Bohm and Hiley on this
score are relevant only to a level of analysis far deeper than would be
appropriate, alas, in almost any contemporary discussion of the foundations
of quantum mechanics.

Bohm's rewriting of Schr\"odinger's equation via variables that seem
interpretable in classical terms does not come without a cost---beyond the
one mentioned above. The most obvious cost is increased complexity:
Schr\"odinger's equation is rather simple, not to mention linear, whereas
the modified Hamilton-Jacobi equation is somewhat complicated, and highly
nonlinear---and still requires the continuity equation for its closure. The
quantum potential itself is neither simple nor natural (even to Bohm it has
seemed ``rather strange and arbitrary'' \cite{bohm}) and it is not very
satisfying to think of the quantum revolution as amounting to the insight
that nature is classical after all, except that there is in nature what
appears to be a rather ad hoc additional force term, the one arising from
the quantum potential.

Moreover, the connection between classical mechanics and Bohmian mechanics
that emerges from the quantum potential is rather misleading. The point is
that Bohmian mechanics is not classical mechanics with an additional force
term. In Bohmian mechanics the velocities are not independent of positions,
as they are classically, but are constrained by the guiding equation $$
v=\nabla S/m.  $$ In classical Hamilton-Jacobi theory we also have this
equation for the velocity, but there the Hamilton-Jacobi function $S$ can
be entirely eliminated and the description in terms of $S$ simplified and
reduced to a finite dimensional description, with basic variables the
positions and momenta of all the particles, given by Hamilton's or
Newton's equations.

It is important to recognize that the guiding equation, expressing the rate
of change $dq/dt$ of the configuration as a specific functional of the wave
function, is not only necessary for Bohmian mechanics but is also
sufficient.  Given any solution $\psi(q,t)$ of Schr\"odinger's equation,
the right hand side of the guiding equation---and hence a first-order
evolution equation for the configuration---is determined for all time. This
equation in turn determines the configuration at any time in terms of the
configuration at any other time.  In particular, since a solution of
Schr\"odinger's equation is determined by the wave function at any initial
time, Bohmian mechanics is entirely deterministic, in the sense that the
initial wave function and configuration completely determine the motion of
the system under consideration.

What concerns us here is the fact that the dynamics for Bohmian mechanics
is thus completely defined by Schr\"odinger's equation together with the
guiding equation, and there is neither need nor room for any further {\it
axioms\/} involving the quantum potential! Thus the quantum potential
should not be regarded as fundamental, and we should not allow it to
obscure, as it all too easily tends to do, the most basic structure
defining Bohmian mechanics.

This is not to say that the quantum potential is of no value. One would
naturally expect the modified Hamilton-Jacobi equation to be convenient,
for example, when considering the classical limit of Bohmian mechanics, in
which the quantum potential is negligible, and Bohm and Hiley have an
excellent presentation of this.  However, in just about all of the other
applications of the quantum potential found in the book of Bohm and Hiley,
the quantum potential itself does not play a genuine role in the analysis;
rather in all such cases it is the wave function and the guiding equation
that are relevant to the analysis.

To my mind, the most serious flaw in the quantum potential formulation of
Bohmian mechanics is that it gives a completely wrong impression of the
lengths to which we must go in order to convert orthodox quantum theory
into something more rational. The quantum potential suggests, and indeed it
has often been stated, that in order to transform Schr\"odinger's equation
into a theory that can account, in what are often called ``realistic''
terms, for quantum phenomena, many of which are dramatically nonlocal, we
must incorporate into the theory a quantum potential of a grossly nonlocal
character.

But note again that Bohmian mechanics incorporates Schr\"odinger's equation
into a rational theory describing the motion of particles merely by adding
a single equation, the guiding equation, a first-order evolution equation
for the configuration. Moreover, the form of the right hand side of this
equation is already suggested by the (pre-Schr\"odinger equation) de
Broglie relation $p=\hbar k$, as well as by the eikonal equation of
classical optics.

Furthermore, if we take Schr\"odinger's equation directly into account---as
of course we should since we seek its rational completion---a form for the
velocity emerges in an almost inevitable manner, indeed via several routes.
Bell's preference is to observe that since standard quantum mechanics
provides us with a probability current $J$ as well as with a probability
density $\rho$, which are classically related (as they are for any dynamics
given by a first-order ordinary differential equation) by $J=\rho v$, it
requires no great imagination to write $$v=J/\rho,$$ which is completely
equivalent to the guiding equation as written above. We should thus not be
surprised to find Bell writing the following about the two-slit experiment:
\bq \noindent Is it not clear from the smallness of the scintillation on
the screen that we have to do with a particle? And is it not clear, from
the diffraction and interference patterns, that the motion of the particle
is directed by a wave? De Broglie showed in detail how the motion of a
particle, passing through just one of two holes in screen, could be
influenced by waves propagating through both holes.    And so influenced
that the particle does not go where the waves cancel out, but is attracted
to where they cooperate. This idea seems to me so natural and simple, to
resolve the wave-particle dilemma  in such a clear and ordinary way, that
it is a great mystery to me that it was so generally ignored. (Bell, page 191)
\eq

My own preference is to proceed in a somewhat different manner, avoiding
any use of probablistic notions even in the motivation for the theory, and
see what symmetry considerations alone might suggest. Then one finds (see
the first reference of \cite{DGZ})
that the simplest choice, compatible with overall Galilean and
time-reversal invariance, for an evolution equation for the configuration
is given by $$v=\frac {\hbar}m\mbox {Im}\frac{\nabla \psi}{\psi},$$ which
also is completely equivalent to the original guiding equation.

It thus seems fair to say that with regard to nonrelativistic quantum
mechanics, the essential departure of Bohmian mechanics from quantum
orthodoxy is merely to insist that particles have positions regardless of
whether or not they are observed. We arrive at the specific evolution
equations of Bohmian mechanics and a rational completion of Schr\"odinger's
equation simply by assigning, in the most obvious way, a role to the wave
function in the evolution of these positions.

\section*{Intuitive Explanation}

In connection with their attitude towards nonlocality, we have seen that
Bohm and Hiley ascribe considerable importance to intuitive explanation, a
rather vague concept that Bell avoids. It is not unreasonable to suppose
that the difference of opinion described above with regard to the quantum
potential originates in a difference with respect to intuitive
explanation, a notion that presumably involves explanation in familiar
terms, for example in terms based on the concepts of classical
mechanics---hence the invocation of the quantum potential.

It hardly seems necessary to remark, however, that physical explanation,
even in a realistic framework, need not be in terms of classical physics.
Moreover, when classical physics was first propounded by Newton, this
theory, invoking as it did action at a distance, did not provide an
explanation in familiar terms. Even less intuitive was Maxwell's
electrodynamics, insofar as it depended upon the reality of the
electromagnetic field. We should recall in this regard the lengths to which
physicists, including Maxwell, were willing to go in trying to provide an
intuitive explanation for this field as some sort of disturbance in a
material substratum to be provided by the Ether. These attempts of course
failed, but even had they not, the success would presumably have been
accompanied by a rather drastic loss of mathematical simplicity.

In the present century fundamental physics has moved sharply away from the
search for such intuitive explanations in favor of explanations having an
air of mathematical simplicity and naturalness, if not inevitability, and
this has led to an astonishing amount of progress. In this regard it must
be emphasized that the problem with orthodox quantum theory is not that it
is unintuitive. Rather the problem is that

\bq \noindent \dots\  conventional formulations of quantum theory, and of
quantum field theory in particular, are unprofessionally vague and
ambiguous. Professional theoretical physicists ought to be able to do
better. Bohm has shown us a way. (Bell, page 173)
\eq
The problem, in other words, is not that orthodox quantum theory fails to be
intuitively formulated, but rather that, with its incoherent babble about
measurement,  it  is not even  well formulated!

\section*{Spin}

This difference of opinion concerning the importance of ``intuitive
explanation'' is perhaps put  in sharpest relief by the respective
approaches of Bell and of Bohm and Hiley to the treatment of spin in
Bohmian mechanics. Spin is the canonical observable having no classical
counterpart, reputed to be impossible to grasp in a nonquantum way. The
source of the difficulty is not so much that spin is quantized in the sense
that its allowable values form a discrete set (for a spin--$\frac12$
particle, $\pm \hbar/2$)---energy too may be quantized in this sense---nor
even precisely that the components of spin in the different directions fail
to commute and so cannot be simultaneously discussed, measured, imagined,
or whatever it is that we are admonished not to do with noncommuting
observables. Rather the difficulty is that there is no ordinary
(nonquantum) quantity which, like the spin observable, is a 3-vector and
which also is such that its components in all possible directions belong to
the same discrete set. The problem, in other words, is that the usual
relationships among the various components of spin are incompatible with
the quantization conditions on the values of these components.

For example, one such relationship is that if $s_x$ and $s_y$ are the
components of spin along the positive $x$ and $y$ directions, then
$(s_x+s_y)/\sqrt2$ is also the component of spin in a certain
direction (namely, the direction bisecting the positive $x$ and the
positive $y$ directions). This implies that the possible values of the
component of spin in this direction must, for the case of spin--$\frac12$,
be either $0$ or $\pm \sqrt2(\hbar/2)$, none of which are permissible
values for such a spin component.  (In fact, insofar as it is relevant to
the issue of hidden variables, von Neumann's theorem on their impossibility
amounts to no more than the preceding rather trivial observation.)

We thus might naturally wonder how Bohmian mechanics manages to cope with
spin. And the impression we will get from Bohm and Hiley is: only with
great difficulty! Bohm and Hiley first propose a model in which the
electron is a spinning body, a proposal that they ultimately abandon,
having ``reached a point where the attempt to account for the acceleration
in terms of various forces no longer provides any useful insight,'' in
favor of a theory that ``still works consistently as long as we have a
suitable guidance condition.''  But even at this point, at which ``there is
no extended particle that is spinning,'' they still find it necessary to
say that ``spin represents only some average property  of the {\it
circulating\/} orbital motion'' (BH, page  221) [my emphasis].

{}From Bell, however, we get an entirely different impression. The guiding
equation $v=J/\rho$ serves as well to define Bohmian mechanics for a
spinor-valued wave function, governed by the Pauli equation, as it does for
a scalar-valued wave function, governed by Schr\"odinger's equation.
Moreover, this guiding equation is also what we find \cite{survey} if we
express the symmetry-based form of the guiding equation, $v=\frac
{\hbar}m\mbox {Im}\frac{\nabla \psi}{\psi},$ in a manner appropriate for a
spinor-valued wave function, namely, $v=\frac {\hbar}m\mbox
{Im}\frac{\psi^*\nabla
\psi}{\psi^*\psi}$. All of the ``mysteries of spin'' are a simple
consequence of the dynamics so defined:

\bq\noindent We have here a picture in which although the wave has two
components [for a single particle], the particle has only position\dots. The
particle does not `spin', although the experimental phenomena associated
with spin are reproduced. Thus the picture \dots\  need not very much resemble
the traditional classical picture that the researcher may, secretly, have
been keeping in mind. The electron need not turn out to be a small spinning
yellow sphere.'' (Bell, page 35)
\eq

\section*{The Quantum Revolution from the Perspective of Bohmian Mechanics}

Bohmian mechanics is more than merely an alternative to the orthodox
Copenhagen interpretation of quantum theory, more than a choice between
equals. After all, orthodox quantum theory, with its invocation of
``measurement'' in a fundamental and irreducible manner, with its appeal to
collapse and to the observer, does not exist as a precise, well-formulated
physical theory. In fact, it could be argued that orthodox quantum theory is
physically vacuous.

This of course raises the question as to how physicists have managed with
such great success to employ orthodox quantum theory---how this theory
could work so well for all practical purposes! The reason for this, I would
argue, is that in using orthodox quantum theory physicists  are thinking in
Bohmian terms---despite the fact that they would claim they are doing
precisely the opposite. Consider, for example, how a physicist thinks about
a scattering experiment: a particle described by a wave packet with more or
less definite momentum and position enters the scattering center and
emerges in a random, though definite, direction, revealed by detectors.  Or
consider, more generally, the prominence of detectors in bringing quantum
experiments to a close---detectors which every physicist, when not in deep
quantum mode, would regard as detecting the particles where these particles
in fact are at! Or consider, in particular, the double-slit experiment, in
which an interference pattern is formed from the accumulated spots marking
the arrivals of the particles at the detector.

Moreover, Bohmian mechanics is not merely a counterexample---to the
assertion that quantum phenomena demand a fundamentally nondeterministic
and subjective description---cooked up to do the job. It has a deep
mathematical and physical integrity. In addition, it provides at least the
beginnings of an answer to the question, ``What exactly is the quantum
revolution?'' I believe that this involves a transition from Newtonian
physics, second-order physics, in which acceleration and forces play a
fundamental role, to first-order physics, in which it is the velocities,
the rates of change of position, that are fundamental in that they are
specified by the theory in a reasonably simple manner. Note that the very
possibility of such a theory, a relativistic (Galilean relativity for the
nonrelativistic case) Aristotelian dynamics as it were, is quite
surprising.

Bohmian mechanics is the natural embedding of Schr\"odinger's
equation---which equation is the common part of almost all interpretations
of quantum theory, however different they may otherwise be, from the
Copenhagen interpretation to the many-worlds interpretation---into a
physical theory. It emerges if one merely insists that the Schr\"odinger
wave function be relevant to the motion of particles. (And notice that if
we are to have a clear physical theory at all, the wave function had better be
relevant to the behavior of {\it something\/} of clear physical
significance!) In other words, Bohmian mechanics arises from
Schr\"odinger's equation when (perhaps naively) we insist upon the simplest
ontology---particles described by their positions---and seek a natural
evolution for this ontology.

All the mysteries of quantum theory find a compelling explanation in
Bohmian mechanics---in the obvious ontology evolving in the obvious way! To
understand how this can be, as well as to develop some appreciation for the
many very difficult remaining open problems in the foundations of quantum
theory, such as that of the tension between nonlocality and Lorentz
invariance, the books of Bell and of Bohm and Hiley are indispensable. Read
them---first Bell, then Bohm and Hiley, and {\it then Bell again!\/} You
will be richly rewarded for your efforts. If you don't quite develop a
clear conception of what the quantum revolution in fact is, you will at
least be confident about what it is not!

\section*{Acknowledgements}
I am grateful to Martin Daumer for his pointed comments and to Rebecca
Goldstein for her instructive observations.  Jean Bricmont deserves double
thanks. This work was supported in part by NSF Grant No. DMS--9305930.

\end{document}